# Quantitative optical coherence tomography reveals rod photoreceptor degeneration in early diabetic retinopathy

David Le BS[1], Taeyoon Son PhD[1], Jennifer I. Lim MD[2], and Xincheng Yao PhD[1,2,*]

[1]Department of Biomedical Engineering, University of Illinois at Chicago, Chicago, IL 60607, USA

[2]Department of Ophthalmology and Visual Sciences, University of Illinois at Chicago, Chicago, IL 60612, USA

[*]Corresponding author: Xincheng Yao, PhD

Richard & Loan Hill Professor, Department of Biomedical Engineering (MC 563)

University of Illinois at Chicago (UIC)

Clinical Sciences North, Suite W103, Room 164D

820 South Wood Street, Chicago, IL 60612

Tel: (312)413-2016; Fax: (312)996-4644; Email: xcy@uic.edu.

**Funding**

National Eye Institute (R01 EY030842, R01 EY030101, R01 EY023522, R01 EY029673, P30 EY001792); Research to Prevent Blindness; Richard and Loan Hill Endowment; T32 Institutional Training Grant for a training program in the biology and translational research on Alzheimer's disease and related dementias (T32AG057468).

**Summary Statement:** Quantitative optical coherence tomography features were used to explore photoreceptor changes in early diabetic retinopathy. The inner segment ellipsoid and retinal pigment epithelium intensity ratio was the most sensitive parameter, and the perifovea was the most sensitive region, which suggests rod abnormalities in early diabetic retinopathy.

# Abstract

**Purpose:** This study is to test the feasibility of optical coherence tomography (OCT) detection of photoreceptor abnormality and to verify the photoreceptor abnormality is rod predominated in early DR .

**Methods:** OCT images were acquired from normal eyes, diabetic eyes with no diabetic retinopathy (NoDR) and mild DR. Quantitative features, including length features quantifying band distances and reflectance intensity features among the external limiting membrane (ELM), inner segment ellipsoid (ISe), interdigitation zone (IZ) and retinal pigment epithelium (RPE) were determined. Comparative OCT analysis of central fovea, parafovea and perifovea were implemented to verify the photoreceptor abnormality is rod predominated in early DR.

**Results:** Length abnormalities between the ISe and IZ also showed a decreasing trend among cohorts. Reflectance abnormalities of the ELM, IZ and ISe were observed between healthy, NoDR, and Mild DR eyes. The normalized ISe/RPE intensity ratio revealed a significant decreasing trend in the perifovea, but no detectable difference in central fovea.

**Conclusions:** Quantitative OCT analysis consistently revealed outer retina, i.e., photoreceptor, changes in diabetic patients with NoDR and mild DR. Comparative analysis of central fovea, parafovea and perifovea confirmed the photoreceptor abnormality is rod predominated in early DR.

# Introduction

Diabetic retinopathy (DR) is the leading cause of blindness in working age adults and is predicted to significantly increase in prevalence worldwide [1]. The number of people affected by diabetes mellitus (DM) is predicted to reach 552 million [2] and nearly 45% of DM patients may develop DR associated vision impairments [3]. Typically, the early stages of DR progress asymptomatically until the patient's vision is affected, however by this time the condition may be irreversible [4]. Therefore, early detection of DR is of the utmost importance to enable prompt treatment to prevent vision loss.

DR encompasses both retinal vascular and neural aspects. Previous studies have demonstrated the close correlation between retinal vascular abnormalities and DR severity [5, 6], and quantitative vascular features of non-proliferative DR (NPDR) have been validated for computer aided DR staging [7, 8]. Recent studies have also reported early retinal neurodegeneration in DR [9-11]. Moreover, electrophysiological features have provided evidence for photoreceptor changes in patients with diabetic retinopathy (DR) [12, 13], suggesting that outer retinal alterations may be observed in optical coherence tomography (OCT).

OCT has enabled depth-resolved visualization of outer retinal changes, especially the external limiting membrane (ELM), photoreceptor inner segment ellipsoid (ISe), and retinal pigment epithelium (RPE), as representative of photoreceptor status. ISe function is closely related to visual acuity (VA) in retinal degenerative diseases [14, 15]. The ELM is also considered to have better reproducibility as another marker of photoreceptor integrity and could be an important predictor of visual function in DR [16]. Although previous studies have revealed the outer retinal features as a DR predictor, these studies were limited to qualitative evaluation of ELM or ISe disruption, measuring outer retinal length changes and no consideration for early DR [17]. We hypothesize that early subtle photoreceptor abnormalities can be quantified using OCT feature analysis. In this study, we evaluate outer retinal band features, length, and reflectance, to reveal early photoreceptor changes in early stages of DR.

## Methods

This is a cross-sectional OCT study evaluating DR biomarkers in patients with diabetes mellitus. Patients with a history of no DR (NoDR) or mild NPDR were recruited from the University of Illinois at Chicago (UIC) Retinal Clinic. This study was conducted in accordance with the ethical standards stated in the Declaration of Helsinki and approved by the institutional review board of the University of Illinois at Chicago. The inclusion criteria included subjects 18 years of age or older with a diagnosis of Type II diabetes mellitus. Exclusion criteria included presence of macular edema, NPDR higher than mild DR prior vitrectomy surgery, history of other ocular diseases other than cataracts or mild refractive error, and ungradable OCT images. All patients underwent a complete anterior segment slit lamp examination and dilated ophthalmoscopy using both the biomicroscope and indirect ophthalmoscopy. The patients were classified as having NoDR or mild NPDR according to the Early Treatment Diabetic Retinopathy Study staging system (ETDRS) by a retina specialist [18].

All patients underwent OCT imaging using spectral-domain OCT (ANGIOVUE spectral domain OCTA system; Optovue, Fremont, CA), with a wavelength of 850 nm, 70-kHz A-scan rate, an axial and lateral resolution of 5 and 15 µm, respectively $6 \times 6$ mm volumetric scan centered at the macula, for a total of either 304 or 400 B-scans, were acquired. All the images were quantitatively examined, and OCT images with severe motion or signal loss were also excluded. OCT volumes were exported into a custom-developed MATLAB (Mathworks, Natick, MA) software for further outer retinal analysis.

**Quantitative OCT Analysis**

Retinal regions for analysis were selected using OCT B-scans centered at the fovea, selection of points between 1.25 and 2.75 mm and 2.5 and 5.5 mm away from the center of the fovea were defined as parafoveal and perifoveal area, respectively (Fig. 1(a)).

A-lines were adjusted from selected areas to match each retinal layer in same horizontal position and averaged, inner limiting membrane (ILM), external limiting membrane (ELM), inner segment ellipsoid (ISe), interdigitation zone (IZ), retinal pigment epithelium (RPE) peaks, and the first and second hyporeflective troughs (T1 and T2) were manually detected (Fig. 1(b)). In this study, we measured six photoreceptor lengths. $L_{12}$, $L_{23}$, $L_{34}$, $L_{13}$, $D_{T1}$, and $D_{T2}$. Where $L_{12}$ quantifies the distance between the ELM to ISe, $L_{23}$ measures the distance from the ISe to IZ, $L_{34}$ measures the distance from IZ to RPE, and $L_{13}$ measures the distance from ELM to IZ. $D_{T1}$ and $D_{T2}$ measures the distance from the ELM to the first and second hyporeflective troughs, respectively. Concurrently, the reflectance intensities of the ELM, ISe, IZ, and RPE were measured from the parafovea and perifovea retina. To reduce noise, the reflectance values were normalized to the inner plexiform layer (IPL) intensity. Additionally, we evaluated the ratio between the ISe and RPE intensity. At the central fovea, the boundary of the IPL cannot be determined, therefore intensity features at the central retina were not analyzed. Each feature was sampled from 10 adjacent A-lines in each retinal region, central fovea, parafovea and perifovea, and the average value was reported.

**Statistical analyses**

Statistical analysis was performed using MATLAB (Mathworks, Natick, MA, USA). One-way ANOVA test was used to compare difference of the mean values of the parameters among different groups. For each feature and each cohort, a post-hoc test using unpaired Student's t-test was performed for one versus one comparison. A *P* value of <0.05 was considered statistically significant.

## Results

The image database used in this study included 14 control subjects (21 eyes), 31 diabetic patients (20 NoDR eyes and 21 Mild NPDR eyes), staged according to the ETDRS staging system [18]. No statistically significant differences were observed among the controls, and diabetic eyes with respect to age, sex, hypertension, or duration of diabetes (analysis of variance [ANOVA], *P* = 0.69, chi-square test, P = 0.85). Furthermore, no significance in hypertension or insulin dependence among the diabetic groups was observed. A summary of the subject demographics used in this study are presented in Table 1.

In this study we quantified 11 outer retinal features, namely 6 length features, $L_{12}$, $L_{23}$, $L_{34}$, $L_{13}$, $D_{T1}$ and $D_{T2}$, and 5 intensity features of the ELM, ISe, IZ, RPE, and ISe/RPE ratio. Examples of averaged A-line intensity plots in the central fovea, parafovea and perifovea for healthy controls, NoDR and Mild NPDR eyes are illustrated in Figure 2. Qualitative observations of the example A-line intensity plots suggest an overall decreasing ISe intensity trend can be observed from control to Mild NPDR eyes, whereas increasing in RPE intensity. Qualitative observations of length changes between cohorts did not reveal a discernable trend, therefore quantitative feature analysis of length and intensity was performed.

For the length features, significant differences were observed for the $L_{2-3}$ feature between mild NPDR and control eyes (Student's t-test, $p<0.013$), and mild NPDR and NoDR eyes (Student's t-test, $p<0.031$) in the parafovea area. Overall, the decreasing trend from no retinopathy to retinopathy can be seen in all three retinal eccentricities. The other length features did not reveal consistent statistically significant differences. All photoreceptor length features are summarized in table 2.

For the intensity features, the ELM intensity revealed statistically significant differences were observed in the perifovea region, however the overall trend with stage progression is neutral. The IZ intensity revealed statistically significant differences between cohorts in the perifovea region, with an overall decreasing trend. The ISe intensity features revealed statistically significant differences between mild NPDR versus control eyes (Student's t-test, $p<0.001$), and mild NPDR versus NoDR eyes (Student's t-test, $p<0.05$). Overall, the trend for ISe intensity decreases with disease progression. Whereas RPE intensity features revealed an increasing trend with disease progression. Since we observed an opposing trend between ISe and RPE intensity, to further highlight differences we took the ISe/RPE intensity ratio. The ISe/RPE ratio revealed a clear decreasing trend with disease progression. There were statistically significant differences between all stages in the perifovea region (Student's t-test, $p<0.05$). However, there was no detectable difference in the central fovea. All intensity features are summarized in table 3.

## Discussion

In summary, we evaluated outer retina alternations in early-stage DR using OCT length and intensity features using clinical OCT. Namely we measured the lengths of the hyperreflective bands, $L_{1-2}$, $L_{2-3}$, $L_{3-4}$, $L_{1-3}$, and hypo-reflective troughs, $D_{T1}$, and $D_{T2}$, and the intensity of the hyperreflective bands in the central fovea, parafovea, and perifovea. The result of this study suggests that there may be metabolic abnormalities that occur in early DR, due to length and intensity changes associated with the ISe.

Previous OCT studies have primarily measured retinal length changes in DR. Goebel et. al., evaluated retinal length in the parafovea retina and reported significant increase in the retinal length of diabetic eyes compared to healthy controls [17]. However, that study was limited by absence of inclusion of analysis of DR stages, particularly NoDR and Mild NPDR. In contrast, a study by Vujosevic et. al., which performed retinal length features for healthy controls, NoDR, and NPDR groups reported no statistical differences in the outer retinal length [19]. Similarly, Dimitrova et. al. compared the differences in the outer retinal length of healthy controls and NoDR cohorts and reported no significant differences among the groups [20]. A recent study by McAnany et. al. performed an analysis of the outer retinal length for healthy controls, NoDR, and mild DR cohorts [21]. In these studies, there may be discrepancies in how they measure the outer retinal lengths, for instance in McAnany et. al., the outer retina length was determined as the boundary between the inner nuclear layer and the outer plexiform layer, which may dilute the subtle length changes of the individual retinal bands. Furthermore, a recent study by Yao et. al., the retinal eccentricities affects the different retinal band lengths [22]. Therefore, in this study we measured the lengths between each band in the photoreceptor layer, for the central fovea, parafovea, and perifovea. The observation in this study suggests that there is a subtle decrease in length with DR stage progression between the 2nd and 3rd band, the location of the ISe to IZ.

Clinically, DR has been defined as a microvascular disease. Therefore, recent endeavors for early detection and objective classification of DR have primarily explored the retinal vasculature for the detection of early-stage DR using OCTA [7, 8, 20, 23]. However, there is a growing body of evidence that suggests that the photoreceptor cells play a role in the development of early stages of DR [24]. The photoreceptor layer is known to be the most metabolically demanding section of the cells in the retina, consuming more than 75% oxygen of the retina [25, 26]. Oxidative stress has been known to play a central role in the complications of diabetes [27]. In relation to the cellular level, mitochondrial DNA is highly susceptible to oxidative damage [28]. Previous studies has shown that elevated levels of superoxides in the retina induces mitochondrial dysfunction [29]. Therefore, to assess retina function, it is crucial to quantify the intensity changes of the photoreceptor layer. The photoreceptor contains more than 75% of the retinal mitochondria, and the ISe is the main location of the photoreceptor mitochondria [26]. Therefore, the ISe may represent the metabolic activities of the photoreceptors. While both the ELM and RPE are not neuronal cells, studies have shown that they do indicate the integrity of the photoreceptor [16]. Recent studies have suggested alterations of the RPE in diabetes, in particular electron microscopy experiments have reported ultrastructural changes in early stage DR [30].

In this study, we assessed the reflectance intensity changes of the ELM, ISe, IZ and RPE, and derived the ISe/RPE intensity ratio. It was observed that the ISe intensity decreases with increased progression of early DR. A study by Toprak et. al., similarly, observed a decrease in ISe intensity in mild NPDR as compared to healthy eyes [10]. Furthermore, we observed an increasing trend for the RPE intensity with increasing severity. Therefore, the relationship between two features with polarizing trends, the ISe/RPE intensity ratio, may enhance the subtle ISe and RPE abnormalities that occur in early DR. Significant differences in the ISe/RPE intensity ratio was observed in the parafovea and perifovea regions. However, there were no significant differences in the central fovea. Since the parafovea and perifovea regions are primarily rod dominated, the observation in this study suggests rod abnormalities in early DR.

This study did have a few limitations, namely our sample size was modest for each cohort, and all the OCT data were acquired from one imaging device (ANGIOVUE spectral domain OCTA system) in a single location. For future study, we plan to expand the population and evaluate OCT data from different devices (e.g., Spectralis, Cirrus, etc.).

In conclusion, quantitative OCT analysis consistently revealed photoreceptor abnormality in diabetic patients with NoDR and mild NPDR. The normalized ISe/RPE intensity ratio of perifoveal OCT is the most sensitive feature to differentiate all three cohorts. Comparative analysis of central fovea, parafovea and perifovea confirmed the photoreceptor abnormality is rod predominated in early DR.

## Disclosures

The authors declare that there are no conflicts of interest related to this article.

**Figure legends:**

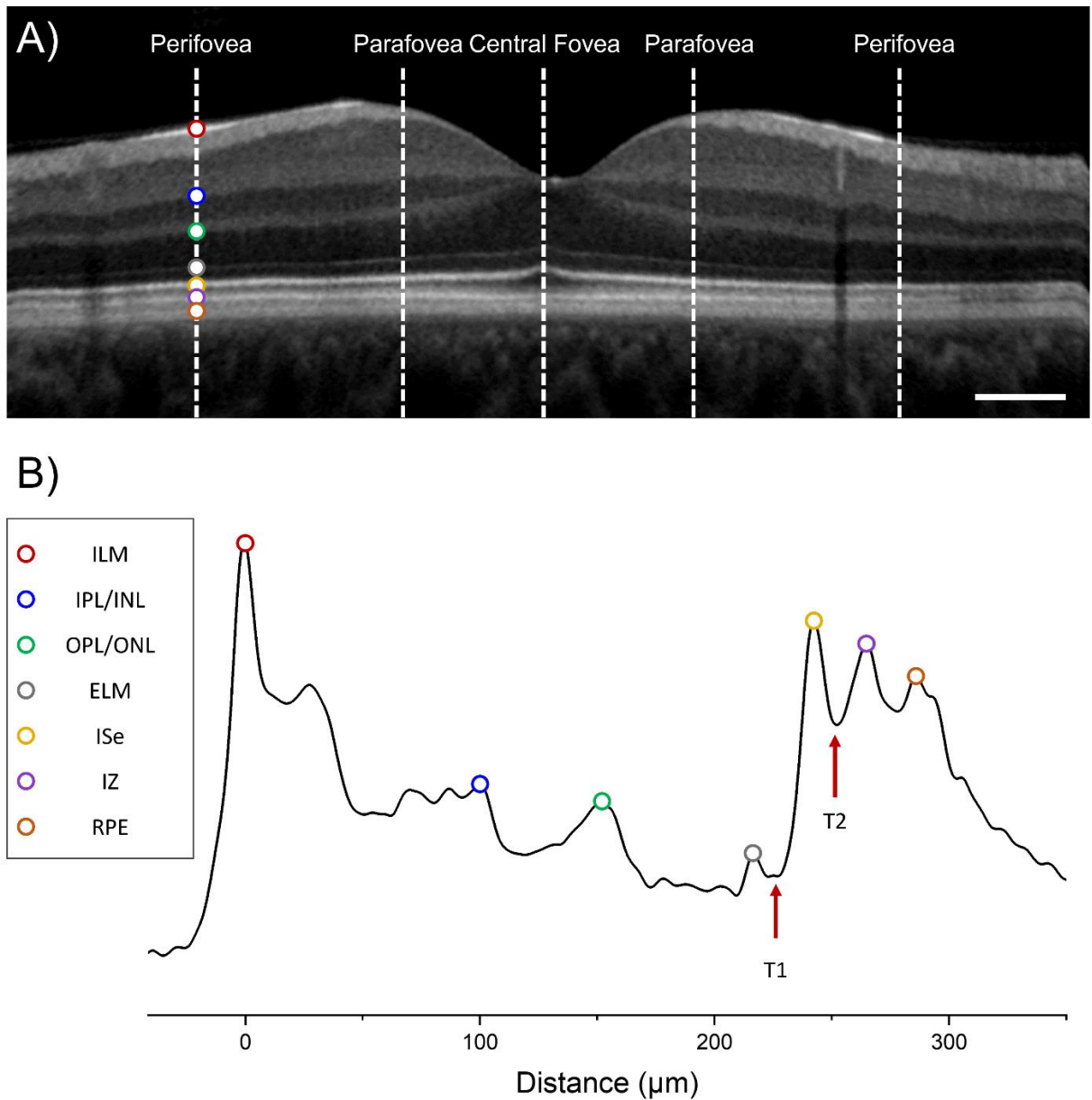

Figure 1. A) Representative OCT B-scan of healthy control subject. B) Representative averaged A-line indicating individual retinal layer location. Outer retina features were performed by peak and trough location detection. ILM: inner limiting membrane; IPL: inner plexiform layer; INL: inner nuclear layer; OPL: outer plexiform layer; ONL: outer nuclear layer; ELM: external limiting membrane; ISe: inner segment ellipsoid; IZ: interdigitation zone; RPE: retinal pigment epithelium; T1: first hyporeflective trough, T2: second hyporeflective trough.

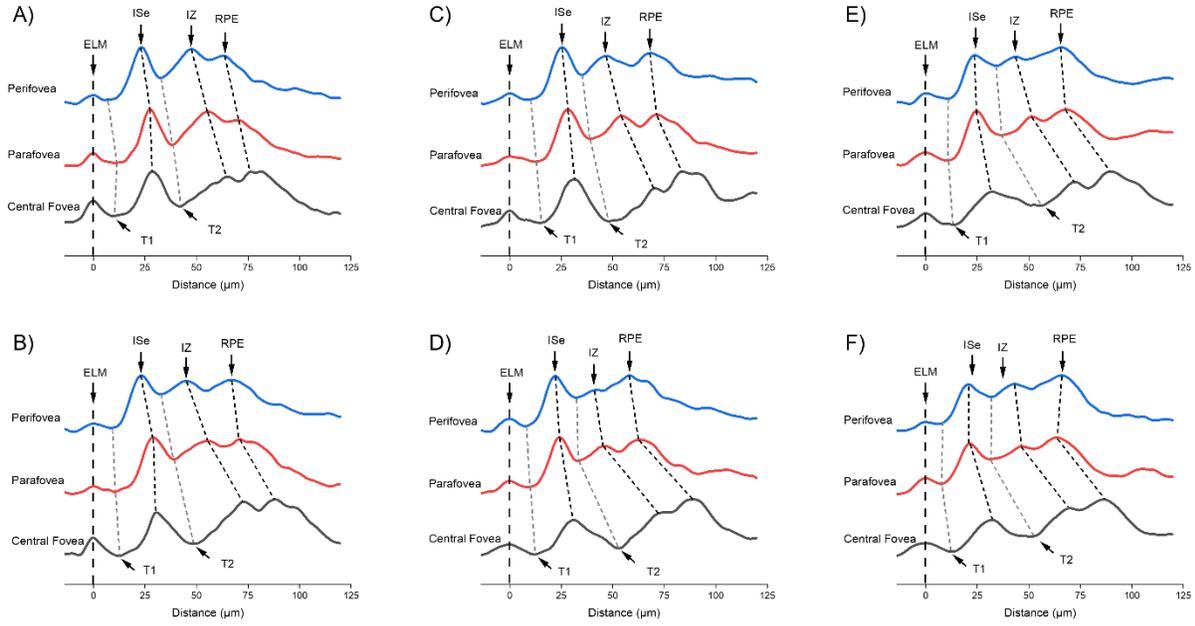

Figure 2. Representative reflectance profiles at central fovea, parafoveal, and perifoveal regions. A-B are from control subjects, C-D are from NoDR patients, and E-F are from Mild NPDR patients. ELM: external limiting membrane; ISe: inner segment ellipsoid; IZ: interdigitation zone; RPE: retinal pigment epithelium, T1: hyporeflective trough 1; T2: hyporeflective trough 2.

Table 1. Demographics of healthy and diabetic subjects.

| | Control | NoDR | Mild NPDR |
|---|---|---|---|
| No. of subjects (n) | 14 | 14 | 17 |
| Age (years) | 58.07 ± 13.43 | 62.27 ± 10.51 | 59.53 ± 11.87 |
| Age range | 38-80 | 47-80 | 24-74 |
| Gender (male) | 8 | 5 | 7 |
| Duration of diabetes (years) | - | 8.89 ± 5.04 | 16.73 ± 4.82 |
| Diabetes type | - | Type II | Type II |
| Insulin dependence (Y/N) | - | 3/11 | 5/12 |
| HbA1c (%) | - | 7.6 ± 2.1 | 8.3 ± 2.4 |
| HTN prevalence, % | 10 | 72 | 70 |

HbA1C, glycated hemoglobin; HTN, hypertension

Table 2. Quantitative analysis of outer retinal length features for different eccentricities and among healthy, NoDR, and mild NPDR cohorts.

| Features | Eccentricity | Control (I) | NoDR (II) | Mild (III) | I vs II | I vs III | II vs III | ANOVA |
|---|---|---|---|---|---|---|---|---|
| $L_{1-2}$ (µm) | Central Fovea | 30.6 ± 2.1 | 32.1 ± 4.1 | 31.7 ± 3.6 | 0.229 | 0.253 | 0.756 | 0.398 |
| | Parafovea | 25.9 ± 2.4 | 24.9 ± 2.5 | 24.6 ± 2.6 | 0.247 | 0.106 | 0.746 | 0.236 |
| | Perifovea | 23.1 ± 2.7 | 22.1 ± 2.8 | 22.9 ± 2.2 | 0.269 | 0.710 | 0.385 | 0.473 |
| $L_{2-3}$ (µm) | Central Fovea | 39.8 ± 3.1 | 39.4 ± 4.5 | 36.6 ± 3.8 | 0.800 | **0.005** | 0.062 | **0.019** |
| | Parafovea | 26.0 ± 3.1 | 25.5 ± 2.3 | 22.7 ± 4.9 | 0.583 | **0.013** | **0.031** | **0.014** |
| | Perifovea | 22.7 ± 2.6 | 23.4 ± 2.9 | 20.7 ± 3.8 | 0.512 | 0.054 | **0.027** | 0.060 |
| $L_{3-4}$ (µm) | Central Fovea | 13.4 ± 3.1 | 13.1 ± 1.9 | 15.3 ± 2.8 | 0.689 | 0.054 | **0.009** | **0.035** |
| | Parafovea | 16.7 ± 3.6 | 14.8 ± 2.5 | 16.5 ± 3.9 | 0.071 | 0.806 | 0.136 | 0.251 |
| | Perifovea | 17.6 ± 2.9 | 15.2 ± 3.2 | 16.7 ± 5.2 | **0.036** | 0.511 | 0.297 | 0.235 |
| $L_{1-3}$ (µm) | Central Fovea | 70.4 ± 3.8 | 71.6 ± 3.5 | 68.3 ± 4.9 | 0.375 | 0.131 | **0.028** | 0.070 |
| | Parafovea | 51.9 ± 4.5 | 50.4 ± 3.8 | 47.3 ± 5.4 | 0.292 | **0.005** | 0.057 | **0.009** |
| | Perifovea | 45.9 ± 3.6 | 45.4 ± 4.5 | 43.6 ± 4.5 | 0.769 | 0.076 | 0.244 | 0.189 |
| $D_{T1}$ (µm) | Central Fovea | 12.9 ± 2.5 | 13.9 ± 3.0 | 13.3 ± 3.1 | 0.329 | 0.702 | 0.546 | 0.623 |
| | Parafovea | 10.7 ± 2.6 | 10.1 ± 1.9 | 10.3 ± 2.2 | 0.406 | 0.571 | 0.763 | 0.699 |
| | Perifovea | 8.4 ± 2.4 | 8.1 ± 1.8 | 8.4 ± 2.0 | 0.696 | 0.989 | 0.669 | 0.912 |
| $D_{T2}$ (µm) | Central Fovea | 47.5 ± 2.9 | 49.3 ± 4.7 | 48.6 ± 5.0 | 0.227 | 0.417 | 0.669 | 0.493 |
| | Parafovea | 36.6 ± 3.1 | 35.1 ± 3.0 | 34.4 ± 3.2 | 0.182 | **0.034** | 0.508 | 0.088 |
| | Perifovea | 32.7 ± 2.7 | 31.7 ± 3.3 | 32.0 ± 2.7 | 0.350 | 0.397 | 0.789 | 0.555 |

Table 3. Quantitative analysis of outer retinal intensity-based features for different eccentricities and among healthy, NoDR, and mild NPDR cohorts.

| Features | Eccentricity | Control (I) | NoDR (II) | Mild (III) | I vs II | I vs III | II vs III | ANOVA |
|---|---|---|---|---|---|---|---|---|
| ELM Intensity | Parafovea | 0.863 ± 0.054 | 0.885 ± 0.042 | 0.881 ± 0.029 | 0.177 | 0.186 | 0.739 | 0.246 |
| | Perifovea | 0.841 ± 0.042 | 0.879 ± 0.042 | 0.829 ± 0.037 | **0.013** | 0.348 | **0.001** | **0.002** |
| ISe Intensity | Parafovea | 1.378 ± 0.073 | 1.386 ± 0.085 | 1.336 ± 0.059 | 0.783 | **0.047** | 0.07 | 0.079 |
| | Perifovea | 1.375 ± 0.072 | 1.383 ± 0.065 | 1.292 ± 0.064 | 0.755 | **0.001** | **0.001** | **0.001** |
| IZ Intensity | Parafovea | 1.361 ± 0.092 | 1.405 ± 0.113 | 1.321 ± 0.091 | 0.240 | 0.155 | **0.028** | **0.048** |
| | Perifovea | 1.348 ± 0.072 | 1.371 ± 0.078 | 1.29 ± 0.079 | 0.400 | **0.017** | **0.006** | **0.006** |
| RPE Intensity | Parafovea | 1.336 ± 0.062 | 1.378 ± 0.054 | 1.392 ± 0.061 | **0.044** | **0.006** | 0.482 | **0.012** |
| | Perifovea | 1.339 ± 0.050 | 1.396 ± 0.059 | 1.351 ± 0.072 | **0.007** | 0.547 | 0.054 | **0.030** |
| ISe/RPE Intensity Ratio | Central Fovea | 0.843 ± 0.072 | 0.827 ± 0.046 | 0.845 ± 0.035 | 0.446 | 0.921 | 0.234 | 0.596 |
| | Parafovea | 1.032 ± 0.047 | 1.007 ± 0.059 | 0.961 ± 0.047 | 0.186 | **0.001** | **0.024** | **0.001** |
| | Perifovea | 1.027 ± 0.040 | 0.992 ± 0.051 | 0.957 ± 0.039 | **0.040** | **0.001** | **0.045** | **0.001** |